\documentstyle[preprint,aps]{revtex}
\begin{document}
\tightenlines
\draft
\title{Boundary effects in a random neighbor model of earthquakes}
\author{Stefano Lise$^{1,a}$ and  Attilio L. Stella$^{2,b}$}
\address{
~\\$^1$ INFM -- Istituto Nazionale per la Fisica della Materia, and \\
     SISSA/ISAS -- International School for Advanced Studies, \\
 Via Beirut 2--4, I-34014 Trieste, Italy \\
$^2$ 
INFM--Dipartimento di Fisica e Sezione INFN, \\
Universit\`a di Padova,  35131 Padova, Italy}
\maketitle
\begin{abstract}
We introduce spatial inhomogeneities (boundaries) in a random neighbor 
version of the Olami, Feder and Christensen model 
[Phys. Rev. Lett. {\bf 68}, 1244 (1992)] 
and study the distributions of avalanches starting both from the bulk
and from the boundaries of the system. Because of their clear geophysical
interpretation, two different 
boundary conditions have been considered (named {\em free} and {\em open},
respectively). 
In both cases the bulk distribution is described by 
the exponent $\tau \simeq \frac{3}{2}$.  Boundary distributions are instead 
characterized by two different exponents $\tau ' \simeq \frac{3}{2}$ 
and $\tau ' \simeq \frac{7}{4}$, for free and open boundary conditions, 
respectively.
These exponents indicate that the mean--field behavior of this  model is 
correctly described by a recently proposed  inhomogeneous form of  
critical branching process.\\
\end{abstract}
\vspace{1cm}
\pacs{PACS numbers: 64.60.Lx, 64.60.Ht, 5.40.+j}
\narrowtext

%
%

Some ten years ago Bak,Tang and Wiesenfeld (BTW) introduced the concept of
self--organized criticality (SOC)\cite{bak}. 
The term refers to the tendency of a 
large class of extended dynamical systems to spontaneously organize 
into a dynamical critical state characterized by long range correlations, 
in both space and time. Many models displaying SOC have been introduced and 
used to describe, for instance, sandpiles\cite{bak}, 
earthquakes\cite{olami}  and biological evolution\cite{sneppen}.
The common feature of all these models is the simplicity of the local dynamical
rules, to be contrasted to the global complex structures 
(e.g., fractal) that emerge as a result of the continued local  interactions. 
A model which has recently attracted much attention  in the literature  
is that proposed by Olami, Feder and Christensen 
(OFC)\cite{olami}, to mimic earthquake dynamics. 
This model has revealed to be particularly challenging to 
understand and  has raised many fundamental questions on the very  
nature of SOC\cite{olami,socolar,grassberger,middleton}.

%
%

The OFC model is a coupled--map lattice model, where to each site $(i,j)$  
of a square lattice, is associated a continuous ``energy'', $E_{ij}$, 
initially fixed at a random value in the interval $(0,E_c)$. 
All the energies  are increased  uniformly and simultaneously at the same 
speed,
until one of them reaches the threshold value $E_c$ and
becomes unstable ($E_{ij} \ge E_c$). The uniform driving is then stopped
and an ``earthquake'' starts:
\begin{equation}
\label{dyn}
E_{i,j} \geq E_c  \Rightarrow \left\{ \begin{array}{l}
                                       E_{i,j} \rightarrow 0 \\
                         E_{nn} \rightarrow E_{nn} + \alpha E_{i,j}
                                      \end{array} \right. 
\end{equation}
where $nn$ denotes the set of nearest neighbor sites of $(i,j)$ and 
$\alpha \in [0, \frac{1}{4}]$ is a parameter controlling the level 
of conservation of the dynamics ($\alpha = \frac{1}{4}$ corresponds to the
conservative case).
The ``toppling'' rule (\ref{dyn}) can possibly produce a chain reaction, which 
ends when all sites are stable again ($E_{kl} < E_c$, $\forall kl$). 
The uniform growth then starts again. 
Boundary conditions imply that sites close to boundaries of the system 
have a smaller coordination number (except in the periodic case) and, 
in principle, also a different  value for $\alpha$ ($\alpha _{BC}$).
Open boundary conditions are generally  set. 
In this case $\alpha _{BC} = \alpha$. \\
The absence of a characteristic length scale in the system is reflected
by the behavior of the probability $P(s)$ that an earthquake involves 
$s$ sites.  Indeed, in the stationary state, $P(s) \propto s^{-\tau}$, 
where $\tau$ is a critical exponent.
 
%
%
Contrary to the BTW sandpile model\cite{manna}, the OFC model is 
believed to be self--organized critical even when the dynamics is 
non conserving
($\alpha < \frac{1}{4}$)\cite{olami,socolar,grassberger,middleton}. 
In a sense this is a rather surprising, 
not yet fully understood result, which possibly implies a peculiar  
mechanism leading to  SOC, different from that of conserved models.
Moreover, the avalanche distribution exponent $\tau$ is 
non--universal: 
it seems to depend on the conservation parameter $\alpha$ and 
even on the type of boundary conditions\cite{olami}.\\
One of the peculiarities of this model is the role played by boundaries.
Besides being ``sinks'' for the energy in excess in the system (which,
moreover, can be dissipated even in the bulk, due to the non-conserving
dynamics), they act as inhomogeneities which frustrate the natural tendency
of the model to synchronize. This is believed to be  the
fundamental mechanism producing SOC in this system
\cite{grassberger,middleton}.
Indeed, it has been shown that, with periodic boundary conditions and
for sufficiently small values of the conservation parameter $\alpha$ 
($\alpha \simeq 0.18$),
the system reaches a strictly periodic state, in which
each avalanche involves just one site\cite{socolar,grassberger,middleton}.
For larger values of $\alpha$ the
situation is slightly more complicated with multiple topplings involved 
in a single avalanche,
but the avalanches are still localized and criticality is not observed.
Open boundaries create a discontinuity in an homogeneous system and break
the periodic state it would otherwise reach.
In ref.\cite{middleton}  it was suggested that sites close to the boundaries
start to organize themselves first, building up long range correlations. 
The critical region grows with time, until, in the stationary state, 
it invades the whole lattice.

%
%
Despite the simplicity of local dynamical rules in SOC systems, analytical 
approaches have rarely been successful (for an exception, see e.g. 
\cite{dhar}). This applies in particular to models of the OFC type. 
Most of the results achieved so far have been obtained through 
computer simulations. To overcome this limitation it has sometimes been
useful to consider random neighbor (RN) models\cite{kim,deboer,lise}, 
where each site interacts with randomly chosen sites instead of its 
nearest neighbors on the lattice. 
This considerably simplifies the problem, though retaining some 
essential features of the original model.  Moreover, RN  models
can be seen as mean field descriptions of their  nearest neighbor
counterparts, since spatial correlations are neglected. \\
The RN versions of the BTW sandpile model\cite{bak} and the
Bak and Sneppen (BS) evolutionary model\cite{sneppen} have been solved 
exactly in ref.\cite{kim} and ref.\cite{deboer}, respectively. 
The solutions show that these models become equivalent to a critical 
branching process. 
Accordingly, the avalanche size distribution exponents is $\tau=\frac{3}{2}$.\\
The RN version of the OFC model\cite{lise} is described by
the following toppling rule
\[
E_{i,j} \geq E_c  \Rightarrow \left\{ \begin{array}{l}
                                       E_{i,j} \rightarrow 0 \\
                         E_{rn} \rightarrow E_{rn} + \alpha E_{i,j}
                                      \end{array} \right. \]
where $rn$ stands for $4$ sites chosen randomly in a finite lattice box. 
Open boundary conditions are implemented by requiring that sites
at the box boundary can collect an infinite amount of energy without toppling.
In this model all bulk sites are equivalent and, therefore, no geometrical 
meaning can be attached to boundaries. \\
A numerical investigation of this version of the RN OFC model gave evidence 
that in the stationary state avalanches  are power law distributed in a whole 
range of $\alpha$ values 
($\alpha _c \le \alpha \le 1/4$; $\alpha _c \simeq 0.225$), with an 
exponent $\tau \simeq 3/2$. 
These results have been questioned by some recent works aimed at an
exact analytical control of the RN model\cite{chabanol,broker}. 
By studying a continuum--time equation it was deduced that  
avalanches are localized as soon as $\alpha < 1/4$, although the mean 
avalanche size grows exponentially fast as dissipation tends to zero.
Anyway, as far as the description of boundary effects is concerned,  
this approach is even less satisfactory. 
In fact no ``boundary'' dissipation is explicitly
introduced in ref.\cite{chabanol,broker},  which, strictly, would make the 
system ``explode''  in the conservative limit ($\alpha \rightarrow 1/4$).
 
%
%

The main purpose  of the present work is that of providing a formulation in  
which the OFC model is still of a RN nature, while allowing  
a meaningful distinction between boundary and bulk of the system.  
This formulation, which can be handled numerically, is worth analyzing 
in view  of the important role boundary inhomogeneities are expected to 
play in the nearest neighbor OFC model. Moreover, our model, introducing the
notion of position in the system,
constitutes a substantial improvement of the standard RN one, 
just like a Landau--Ginzburg approach compared to the 
Weiss theory of ferromagnetism.  

%
%
In order to introduce a proper inhomogeneity due to boundary effects,
we have proceeded as follows. We have considered the $2D$--system as 
divided into columns, say from $1$ to $L$ (each column has two 
``boundary'' sites, which are never allowed to discharge).
When a site with energy 
$E > E_c$ in column  $i$ topples, it distributes an energy $\alpha E$ to two 
randomly chosen sites, one in column $i-1$ and another one in column $i+1$. 
Moreover, an energy $\alpha E$ is also received by two randomly chosen  
sites in column $i$.
In this way the notion of position along the direction perpendicular to the
columns acquires a meaning and the effect of the boundaries 
(located at columns $0$ and  $L+1$) can propagate  into the system. 
At the same time, the random choice of sites
within columns should guarantee  that we are dealing with an inhomogeneous 
mean--field model. The set of all columns behaves as a finite chain with 
boundaries,
from which  also bulk behavior can be extracted in the $L\to \infty$ limit.\\
Boundary conditions are determined by setting the 
level of conservation of the dynamics $\alpha$  at  sites close to
the boundaries ($\alpha _{BC}$). Two different possibilities are 
suggested by the Burridge--Knopoff model of earthquakes\cite{burridge}:
a) {\em free} boundary conditions, 
in which $\alpha _{BC}= \frac{\alpha}{1-\alpha}$ and b) 
{\em open} boundary conditions, in which $\alpha _{BC}= \alpha$.
The Burridge--Knopoff model is a driven spring--block model, which can
be directly mapped into  the OFC model\cite{olami}.
It can be schematized as a 2--dimensional network
of blocks interconnected by springs. In addition, all the blocks  are 
subject to an external driving force, which pulls them, and to a static 
friction, which opposes their motion. 
The case of free boundary conditions corresponds to boundary
blocks connected only to blocks belonging to the system.
The case of open boundary conditions, instead, corresponds
to blocks at the boundary   coupled  also to an imaginary  external block
(see ref.\cite{olami} for further details). 
Although both of them are probably not realistic, it is believed  
that more adequate boundary conditions should somehow interpolate between 
these two extreme limits\cite{geo}.
Note moreover that free boundary conditions are more conservative 
than open ones and become strictly conservative for $\alpha = \frac{1}{4}$. 
Below we will show that this distinction leads to
a radical difference in the distributions of avalanches propagating from
the borders of the chain.
%
%

We performed extensive numerical simulations in order to sample bulk
and boundary avalanche size distributions.
The bulk avalanche distribution of the system appear not to affected 
by the choice of boundary conditions. 
Fig. 1(a) and 2(a) report the results in the conservative case 
($\alpha =0.25$) for free and open boundary conditions, respectively. 
Only avalanches starting from the deep interior of the lattice 
have been taken into account in the statistics.
As a matter of fact, bulk behavior is also easily extracted upon, e.g., 
averaging the avalanche distribution over all possible starting columns 
in the lattice.
The estimated exponents are $\tau=1.45 \pm 0.1$ and $\tau=1.5 \pm 0.1$ for 
free and open boundary conditions, respectively.
In both cases they are consistent with the usual mean--field one, i.e. 
$\tau = \frac{3}{2}$.\\
On the contrary, avalanches starting from the borders, i.e. from columns $1$ 
or $L$, are strongly influenced by the choice of boundary conditions.
Boundary avalanches are distributed as $P(s) \propto s^{-\tau '}$, where
$\tau ' = 1.4 \pm 0.1$ for free boundaries (fig. 1(b)) and 
$\tau ' = 1.75 \pm 0.1$ for open boundaries (fig. 2(b)). We have also
checked that boundary conditions interpolating between free and open 
behave as the open case, i.e. $\tau ' \simeq 1.75$. In this 
respect $\tau ' \simeq 1.75$ reflects a more robust behavior of the
model, appropriate to any non zero level of dissipation realized at the
border.\\ 
Finally we have verified that the introduction of bulk dissipation 
($\alpha < 1/4$)
has no apparent effect on bulk and boundary exponents, as long as 
$\alpha  > \alpha _c$, as determined in ref.\cite{lise}. 
 However, it is 
controversial\cite{chabanol,broker} whether for $\alpha < 1/4$
the model should be considered critical. An apparent criticality could
result numerically from finite size effects. 

Our results indicate that the mean--field OFC model with open 
boundary conditions is correctly described by an inhomogeneous branching 
process\cite{gcalda}.
This generalization of the standard branching process has been recently 
proposed and studied as a paradigm for a more complete description of SOC 
models in the mean--field  limit.  
The inhomogeneous branching process takes place in a situation in which 
translation invariance is broken.
An example is given by the geometry of a semi--infinite chain. 
Each site of the chain involved in the process at a given stage of the
process can activate its neighbors (and/or reactivate itself) at the
subsequent stage with well defined probabilities.
The probability of generating a tree (or avalanche) with $s$ individuals 
becomes a function of the position $n=1,\ldots,\infty$ where the tree starts 
along a chain. 
It was shown exactly in ref.\cite{gcalda} that  critical trees (avalanches) 
starting from the  boundary are distributed as $P_1(s) \propto s^{-\tau '}$, 
with  $\tau ' = \frac{7}{4}$, different from the bulk exponent  
$\tau = \frac{3}{2}$. 
Boundary sites were there identified as sites 
with an average number of branchings smaller than $1$\cite{harris}, 
reflecting in this way a sort of ``dissipative'' behavior.
In this respect open boundary conditions in our model correspond to the 
existence of these ``dissipative'' boundaries of
the inhomogeneous branching process. 
On the other hand, if the average number of branchings is maintained equal 
to $1$ for
all sites of the chain  (including the end), it is possible to verify 
numerically that, in the inhomogeneous branching process, the exponent 
$\tau '$
coincides with  the bulk one (i.e. $\tau ' \simeq \frac{3}{2}$)\cite{monte}. 
This case seems to be realized by our model with  free boundary
conditions. Thus, free conditions here seem adequate to keep at its
critical bulk value (i.e. $1$) the average number of descendants for each
generations occurring at the border, in the branching process underlying 
earthquake  dynamics. 

%
%
In conclusion we have shown how it is possible to introduce proper boundary
effects at the level of a RN OFC model.  
The model we have considered can also be interpreted as a $1 D$ 
nearest--neighbor model with ``$\infty$--components''. 
Indeed, the energies in a given column can be thought of as the many
``components'' of an energy vector associated to a single site of a
$1 D$ chain. Since we have always considered very large $L$, our
approach effectively corresponds to an ``$\infty$--components'' limit
of the $1 D$ model.\\
We have verified that, with open boundary conditions, 
boundary avalanches in the OFC model are distributed as 
$P(s) \propto s^{-\frac{7}{4}}$, as predicted by the analysis of the
inhomogeneous branching process.  Free boundary conditions, instead, 
correspond to ``conservative'' boundary conditions in the inhomogeneous 
branching process and imply $\tau ' \simeq \tau \simeq 3/2$. \\
As a prospective for future work, it would be interesting to investigate
whether the mechanism of invasion from the boundaries of the 
``self--organized'' region  (named  ``phase locking'' in 
ref.\cite{middleton}) is actually present also in the random neighbor OFC
model with boundaries. 
 We expect that also in the nearest neighbor OFC model boundary scalings
different from the bulk ones should be observed with appropriate
conditions at the borders. \\
\noindent
We acknowledge Claudio Tebaldi for many useful conversations on the 
subject and for a critical reading of the manuscript. 

\noindent 
$^a$ Email: lise@shannon.sissa.it\\
$^b$ Email  stella@padova.infn.it


\vskip 0.6 truein
\centerline{\large FIGURE CAPTIONS}

\vskip 0.4 truein
\noindent
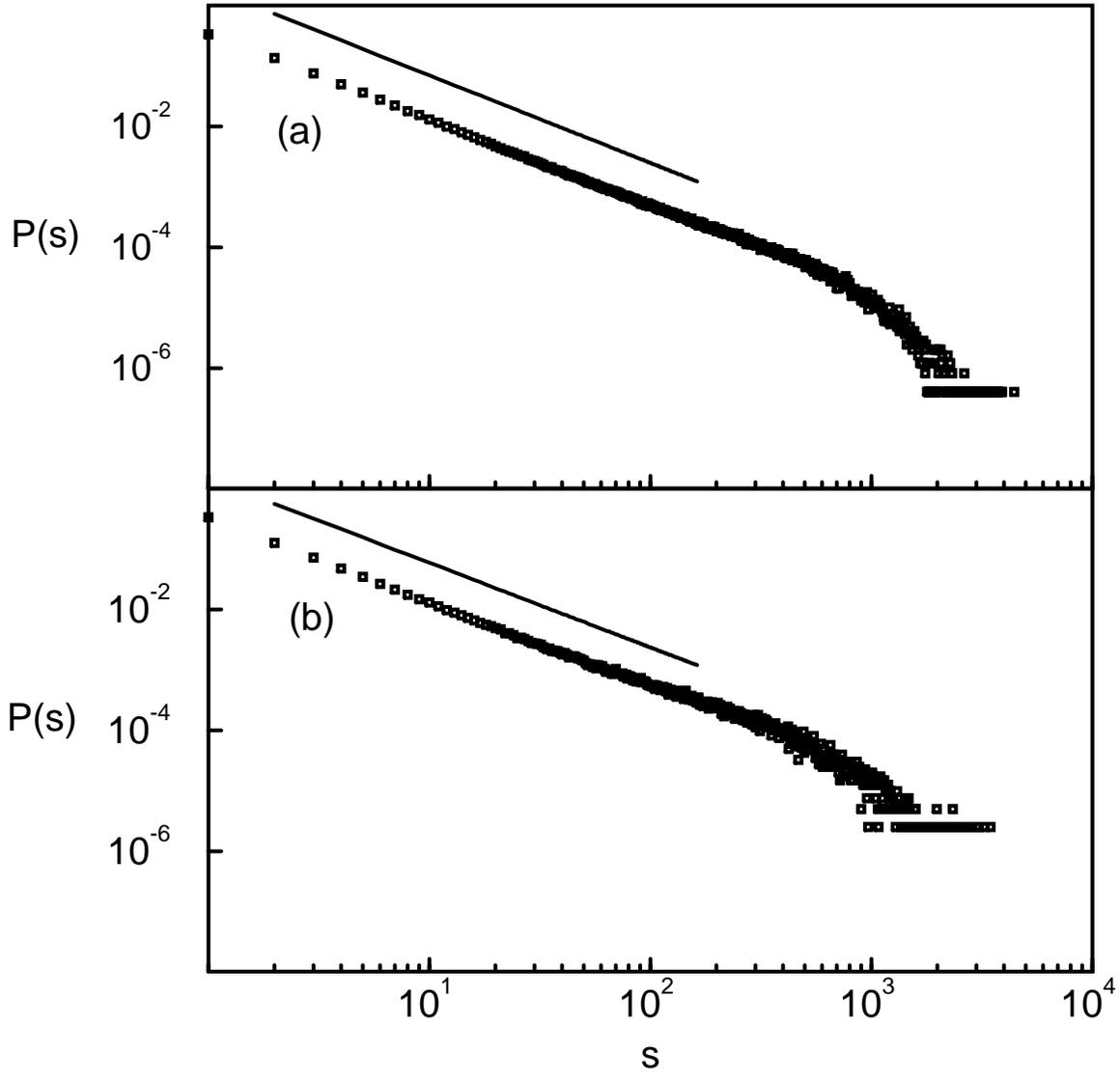
Fig. 1: Simulation results for (a) bulk  and (b) boundary
avalanches in a system with free boundary conditions.
The system size is $L=100$ and the conservation parameter is 
$\alpha=0.25$. The estimated exponents are (a)  $\tau = 1.45 \pm 0.1$
and (b) $\tau '= 1.4 \pm 0.1$.   


\vskip 0.4 truein
\noindent
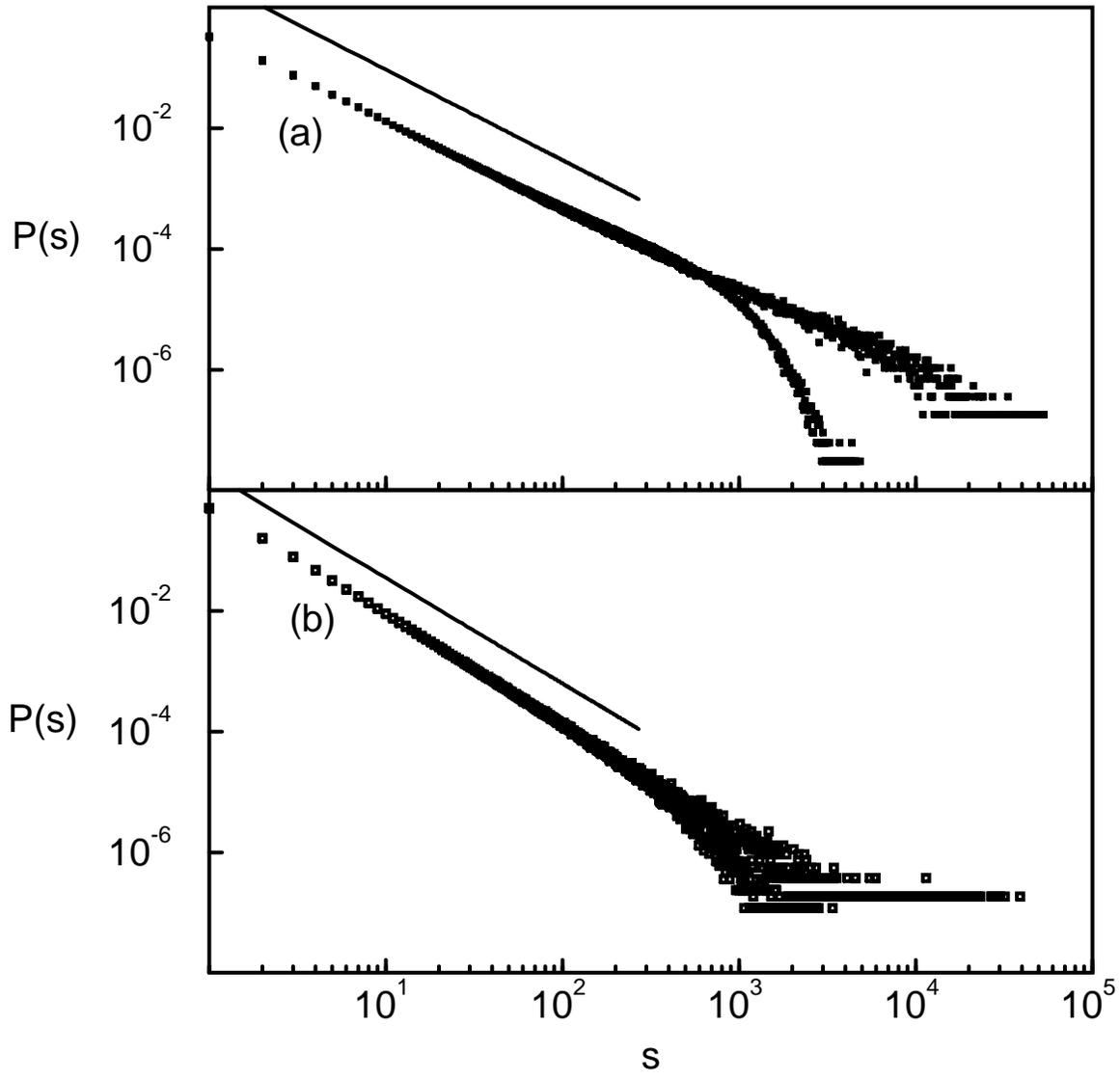
Fig 2:  Results for (a) bulk  and (b) boundary avalanches in a 
system with  open boundary conditions.
System size are  $L=100$, $200$ and conservation parameter is 
$\alpha=0.25$. 
The estimated exponents are (a)  $\tau = 1.5 \pm 0.1$
and (b) $\tau '= 1.75 \pm 0.1$.


\end{document}